\documentclass[a4paper,10pt]{article}
\usepackage{amssymb, amsmath}
\newcommand{\be}{\begin{equation}}
\newcommand{\ee}{\end{equation}}
\newcommand{\br}{{\bf r}}
\newcommand{\bv}{{\bf v}}
\newcommand{\bw}{{\bf w}}
\newcommand{\bk}{{\bf k}}
\newcommand{\bee}{\begin{eqnarray}}
\newcommand{\eee}{\end{eqnarray}}
\title{Three-dimensional Lorentz model in a magnetic field: exact and Chapman-Enskog solutions}
\author{F. Cornu \\
	Laboratoire de Physique Th\'eorique 
	\thanks{Laboratoire associ\'e au Centre National
	de la Recherche Scientifique - UMR 8627}\\
        B\^{a}timent 210, Universit\'e
	Paris-Sud, 91405 Orsay, France
	\and
	J. Piasecki\\ Institute of Theoretical Physics\\ University of Warsaw,
	Ho\.{z}a 69, 00-681 Warsaw, Poland}
\date{ April 6, 2006}

\begin{document}
\maketitle
\begin{abstract}
We derive the exact solution of the Boltzmann kinetic equation for the three-dimensional Lorentz model in the presence of a constant and uniform magnetic field. The velocity distribution of the electrons reduces exponentially fast to its spherically symmetric component. In the long time hydrodynamic limit there remains only the diffusion process governed by an anisotropic diffusion tensor. The systematic way of building the Chapman-Enskog solutions is described. 
\vskip 0.5cm 
{\bf PACS ~:} 05.20.Dd ; 51.10.+y
\vskip 0.5cm 
{\bf KEYWORDS~:} Lorentz model; cyclotron motion; Boltzmann equation; hydrodynamic mode.
\vskip 0.5cm 
{\it Corresponding author ~:} CORNU Fran\c coise, \\
Fax: 33 1 69 15 82 87,
E-mail: cornu@th.u-psud.fr
\end{abstract}

\section{Introduction}

The model proposed by Lorentz in 1905 \cite{Lorentz05} to describe the electrical conductivity in metals has become  an inexhaustible source of results concerning the foundations of the kinetic theory.
Within the model  the electrons are considered as small hard spheres with electric charge $(-e)$ propagating among randomly distributed immobile hard spheres representing the atoms. The electrons do not interact with each other. The elastic electron-atom scattering is at the origin of the electric resistivity of the medium. 

An excellent review of the subject up to 1974 can be found in  \cite{Hauge74}. 
Since then the research keeps bringing new results.  More recent applications of the model, in particular in the theory of dynamical systems have been described in \cite{Dorfman99}. One of the relatively recent discoveries was the non-Markovian character of the evolution in two-dimensions in the presence of a magnetic field perpendicular to the plane of motion. Even in the Grad limit where the fraction of the volume occupied by the scattering atoms is very low the kinetic equation keeps containing  memory effects and does not reduce to the Boltzmann form \cite{BMHH95}. 

We consider here  the three-dimensionsional dynamics where the mean free path of the electron $\lambda $
is inversely proportional to the number density of the scatterers $n$ and to the scattering cross section  
$\pi a^{2}$  with $a$ equal to the sum of the atomic and electronic radii. 
Keeping $\lambda $ fixed while the volume fraction $na^{3}$ approaches zero defines the Grad limit
\begin{equation}
\lim_{\rm Grad} = \left\{  
\begin{array}{ll}
& \lambda  = 1/n\pi a^{2} = const \\  
& na^{3} \to 0 
\end{array}
\right.
\label{G}
\end{equation}
In the absence of external fields the Boltzmann equation applies in the regime (\ref{G}).
Moreover, it can be rigorously solved. This fact has been exploited in \cite{Hauge70} to compare the evolution predicted by the Boltzmann equation for general initial conditions with that  of the special class of Chapman-Enskog solutions involving exclusively hydrodynamic modes. 
 
 Our object is to generalize the analysis of \cite{Hauge70} to the case where the electrons are acted upon by a constant and uniform magnetic  field ${\bf B}$. As we consider the motion in three-dimensions the non-Markovian effects studied in \cite{BMHH95} do not occur  and the Boltzmann equation 
 yields an adequate description in the low density limit (\ref{G}).
The electrons move between collisions under the action of the Lorentz force ${\bf F}$ giving rise to the acceleration
\begin{equation}
\label{2}
\frac{\bf F}{m}= \boldsymbol{\omega}\times \bv = \omega \widehat{\bf B}\times \bv 
\end{equation}
where ${\bf B}=B\widehat{\bf B}$, $\vert\widehat{\bf B}\vert=1$, $m$ is the electronic mass, and
$\omega=eB/m $ is the cyclotron frequency.
 
We denote by $f(\br,\bv,t)$ the probability density for finding the electron at time $t$ at  point $\br$ moving with velocity $\bv$. The Boltzmann equation applied to the Lorentz model (also called Boltzmann-Lorentz equation) reads
\begin{equation}
\label{Boltzmann}
\left[\frac{\partial  }{\partial t} +\bv\cdot\frac{\partial}{\partial \br}+
\left(\boldsymbol{\omega}\times \bv \right) \cdot\frac{\partial}{\partial\bv}\right]f(\br,\bv,t)=\frac{1}{\tau}\left\{\;  [\mathbb{P}f](\br,v,t) - f(\br,\bv,t)\; \right\}
\end{equation}
The collision frequency $1/\tau = v/\lambda$ is constant under the assumed dynamics. Indeed, the speed $v=\vert\bv\vert$ of the electron does not change during the free cyclotron motion and is also conserved by the elastic collisions. 
The gain term  in the right hand side of (\ref{Boltzmann}) represents the effect of isotropic scattering which averages the distribution $f$ over all directions leaving the spherically symmetric part
\begin{equation}
\label{3}
[\mathbb{P}f](\br,v,t)= \int  \frac{d\widehat{\bv}}{4\pi}\, f(\br,\bv,t)
\end{equation}
 $\widehat{\bv}=\bv/v$  represents a point on a unit sphere in the velocity space so that $d\widehat{\bv}$ is the solid angle measure.
 
 It turns out that the initial value problem for the kinetic equation (\ref{Boltzmann}) can be explicitly solved. We begin  by deriving in section 2 the evolution of the velocity distribution
 which permits to calculate the diffusion tensor via a Green-Kubo formula.
The complete exact solution for the distribution $f$ is derived in section 3. Section 4 is devoted to the discussion of the hydrodynamic diffusion process governing the evolution in the long time limit. In section 5 we indicate the method of a systematic  construction of  Chapman-Enskog solutions whose time dependence is entirely defined by the diffusive mode at any time scale. The final section 6 contains discussion and comments.

\section{Velocity distribution and diffusion tensor}

\subsection{Approach to spherical symmetry}
The evolution equation for the velocity distribution
\begin{equation}
 \label{4}
\phi(\bv,t)=\int d\br\, f(\br,\bv,t)
 \end{equation}
 follows by integrating (\ref{Boltzmann})  over the position space. One finds
 \begin{equation}
\label{5}
\left[\frac{\partial  }{\partial t} +
\left(\boldsymbol{\omega}\times \bv \right) \cdot\frac{\partial}{\partial \bv}\right]\phi(\bv,t)=-\frac{1}{\tau}\left[\, \phi(\bv,t)-[\mathbb{P}\phi](v,t)\, \right]
\end{equation}
Applying the projection $\mathbb{P}=\mathbb{P}^2$ to both sides of (\ref{5}) yields the  relation
 \begin{equation}
\label{6}
\frac{\partial  }{\partial t} [\mathbb{P}\phi](v,t)=0
 \end{equation}
The spherical part $[\mathbb{P}\phi]$ of the velocity distribution remains unchanged in the course of time.  This permits to rewrite (\ref{5}) in the form
\begin{equation}
\label{7}
\left[\frac{\partial  }{\partial t} +
\left(\boldsymbol{\omega}\times \bv \right) \cdot\frac{\partial}{\partial \bv}\, \right]\chi (\bv,t)= 0
\end{equation}
with
\begin{equation}
\label{8}
\chi (\bv,t) =   \exp(t/\tau ) [ \; \phi(\bv,t)-[\mathbb{P}\phi ](v,0) \; ] 
\end{equation}
 The solution of (\ref{7}) is straightforward as it consists in propagating $\chi (\bv,t)$ backward in time up to $t=0$ along the free cyclotron  trajectory
 \begin{equation}
 \label{9}
 \bv (-t) =  {\cal R}_{\widehat{\bf B}}(-\omega t)\cdot \bv
=\bv_\parallel +  {\cal R}_{\widehat{\bf B}}(-\omega t)\cdot \bv_\perp
 \end{equation}
where ${\cal R}_{\widehat{\bf B}}\left(\alpha\right)$ is the rotation of angle $\alpha$ around the axis $\widehat{\bf B}$.
 The velocity component parallel to the magnetic field
 $ \bv_\parallel = (\bv\cdot\widehat{\bf B})\widehat{\bf B}$
is not affected by the cyclotron rotation whereas the perpendicular component 
 $\bv_\perp = \bv - \bv_\parallel$  rotates with constant angular velocity $\omega $.
We thus have in general
\[ \chi (\bv,t) = \chi (\bv (-t) , 0 ) \]
which in view of (\ref{8})  yields the solution 
\begin{equation}
\label{vd} 
\phi(\bv,t)= [\mathbb{P}\phi_0](v)+ e^{-t/\tau}\left[\phi_0( {\cal R}_{\widehat{\bf B}}(-\omega t)\cdot \bv)-[\mathbb{P}\phi_0](v)\right]
\end{equation}
with
\begin{equation}
\label{11}
\phi_0(\bv) \equiv \phi(\bv,t=0)
\end{equation}
The velocity distribution becomes spherically symmetric exponentially fast with the relaxation time  $\tau = \lambda/v$. Its projection $[\mathbb{P}\phi_0](v)$ does not change in the course of time and represents the asymptotic equilibrium state.

\subsection{Diffusion tensor from a Green-Kubo formula}

The complete solution of the initial value problem (\ref{vd}) permits to evaluate the conditional probability 
$P(\bv,t\vert \bw ,0)$ for finding the electron with velocity $\bv (t) = \bv $ at time $t$ provided it had velocity $\bv (0) =\bw $ at time $t=0$. One finds
\begin{equation}
\label{13}
 P(\bv,t\vert \bw ,0)  = \frac{\delta (v- w )}{4\pi w^2 }
 +   e^{-t/\tau}\left\{\delta\left[{\cal R}_{\widehat{\bf B}}
(-\omega t)\cdot\bv - \bw \right]
-\frac{\delta ( v- w )}{4\pi w^2 } \right\} 
\end{equation}
The knowledge of $P(\bv,t\vert \bw ,0)$ suffices to determine the diffusion tensor $D_{\mu\nu}$ through  the Green-Kubo formula
\begin{equation}
\label{D}
D_{\mu\nu}  = \int_0^{+\infty}dt\; \langle \bv_\mu(t)\bv_\nu(0)\rangle_{eq} 
 \end{equation}
where the velocity auto-correlation function is given by
\begin{equation}
\label{14}
 \langle \bv_\mu(t)\bv_\nu(0)\rangle_{eq} =
 \int d\bv\int d\bw \, \bv_\mu\bw_\nu P\left(\bv,t\vert \bw,0\right) \phi_{eq}(w)
\end{equation}
Here
            \[  \phi_{eq}(w) = \delta ( w- v_{0} )/4\pi v_{0}^2   \]
denotes the spherically symmetric equilibrium state. 

The $\delta$-distributions greatly facilitate the evaluation of the integral in (\ref{14}) which is readily reduced to
\begin{equation}
\label{15}
\langle \bv_\mu(t)\bv_\nu(0)\rangle_{eq}=
v_0^2 \int  \frac{d\widehat{\bv}}{4\pi}\, \widehat{ \bv}_\mu \left[{\cal R}_{\widehat{\bf B}}
(-\omega t)\cdot \widehat{ \bv}\right]_\nu \, e^{-t/\tau}
= \frac{v_0^2}{3}  \left[{\cal R}_{\widehat{\bf B}}
(\omega t)\right]_{\mu\nu}\, e^{-t/\tau}
\end{equation}
Here $\left[{\cal R}_{\widehat{\bf B}}(\omega t)\right]_{\mu\nu}$ are the elements of the
rotation matrix corresponding to rotation of angle $\omega t $ around the $\widehat{\bf B}$-axis.
The calculation of the diffusion tensor (\ref{D}) is conveniently performed in the coordinate system whose $z$-axis is oriented along the magnetic field. One then finds  the diffusion matrix with diagonal elements
\begin{equation}
\label{Dd}
D_{\perp}=D_{xx}=D_{yy}=\frac{1}{3} v_0^2 \tau\frac{1}{1+\omega^2\tau^2} \; , 
\;\;\;\;
D_{\parallel} = D_{zz}=\frac{1}{3} v_0^2 \tau  
\end{equation}
and the off-diagonal elements
\begin{equation}
\label{Dof}
D_{xy}=-D_{yx}=-\frac{1}{3} v_0^2 \tau\frac{\omega\tau}{1+\omega^2\tau^2}, \;\;\;\;
D_{xz}=D_{yz}=0
\end{equation}
Equations (\ref{Dd}),(\ref{Dof}) show the reducing effect of the magnetic field on the diffusion process  perpendicular to the magnetic field. The fact that the corresponding diffusion coefficient  $D_{\perp}$ decays at large values of the field as $(\omega \tau )^{-2}$ has been well established  in  plasma physics \cite{Lifszic}. The formulae (\ref{Dd}), (\ref{Dof}) have been also derived with the use of the appropriate Langevin equation
\cite{Balescu}. The physical content of expressions (\ref{Dd}) and (\ref{Dof}) is discussed in detail therein.

Our main object in this paper is to exhibit the time scale separation in the evolution toward equilibrium. In contradistinction to the rapid relaxation (\ref{vd}) of velocities, there also appears a by far slower hydrodynamic spreading of the electrons in the position space governed by the diffusion tensor.

\section{Solution of the Boltzmann equation}

On the left hand side of the Boltzmann equation (\ref{Boltzmann}) we find the operator
\begin{equation}
\label{16}
 \left[\frac{\partial  }{\partial t} +\bv\cdot\frac{\partial}{\partial \br}+
\left(\boldsymbol{\omega}\times \bv \right) \cdot\frac{\partial}{\partial\bv}\right] 
\end{equation}
which generates collisionless trajectories in the magnetic field.  The backward in time motion starting from the phase space point (\br,\bv) reads
\bee
\label{17}
\br (-t) & = & \br - \bv_\parallel t -\frac{1}{\omega} {\cal R}_{\widehat{\bf B}}\left(\frac{\pi}{2} \right)\cdot [\, \bv_{\perp} (-t) - \bv_{\perp} \, ]  \nonumber \\
\bv (-t) & = &  {\cal R}_{\widehat{\bf B}}(-\omega t)\cdot \bv=\bv_\parallel +  {\cal R}_{\widehat{\bf B}}(-\omega t)\cdot \bv_\perp
\eee
(see  equation (\ref{9})). The operator (\ref{16}) vanishes when applied to the trajectory (\ref{17}) which permits to rewrite (\ref{Boltzmann}) in the integral form
\begin{equation}
\label{IntegralB}
f(\br,\bv,t)=
e^{-t/\tau}f_0(\br(-t),\bv(-t))+\frac{1}{\tau}\int_0^t dt'
e^{-(t-t')/\tau}[\mathbb{P}f](\br(-(t-t')),v,t')
\end{equation}
with $f_0(\br,\bv) = f(\br,\bv,t=0)$.
Applying to (\ref{IntegralB}) the Fourier transformation 
\begin{equation}
\label{FT}
\hat{f}(\bk,\bv,t)=\int d\br \, e^{-i\bk\cdot\br} f(\br,\bv,t)
\end{equation}
we find
\begin{equation}
\label{FTB}
\hat{f}(\bk,\bv,t)=
\exp\{-t/\tau-i\bk\cdot \left[\bv_\parallel t +{\bf x}(\bv_\perp, t)\right]\}\hat{f_0}
(\bk,{\cal R}_{\widehat{\bf B}}(-\omega t)\cdot \bv)
\end{equation}
\[ + \frac{1}{\tau}\int_0^t dt'
\exp\{-(t-t')/\tau-i\bk\cdot  \left[\bv_\parallel (t-t') +{\bf x}(\bv_\perp, (t-t'))\right]\}[\mathbb{P}\hat{f}](\bk,v,t') \]
where the shorthand notation
\begin{equation}
\label{18}
{\bf x}(\bv_\perp, t)=\frac{1}{\omega} {\cal R}_{\widehat{\bf B}}\left(\frac{\pi}{2} \right)\cdot 
[\, \bv_{\perp} (-t) - \bv_{\perp} ]
\end{equation}
has been introduced. 

The way to the solution consists now in using the fact that the Laplace transformation
\begin{equation}
\label{LT}
\tilde{f}(\bk,\bv,z)=\int_0^{+\infty} dt \, e^{-zt}\hat{f}(\bk,\bv,t)
\end{equation}
factorizes the convolution in time present in (\ref{FTB}).
The integral form (\ref{IntegralB}) of the Boltzmann equation for the Fourier-Laplace transform $\tilde{f}$
takes the form
\begin{equation}
\label{LFTB}
\tilde{f}(\bk,\bv,z)=\left[\mathbb{L}\hat{f_0}\right](\bk,\bv,z) 
+\psi(\bk,\bv,z){[\mathbb{P}\tilde{f}]}(\bk,v,z) 
\end{equation}
where the direct effect of the initial state is contained in
\begin{equation}
\label{19}
\left[\mathbb{L}\hat{f_0}\right](\bk,\bv,z)=
\end{equation}
\[ \int_0^{+\infty} dt\, \exp\left\{-\left( z+\frac{1}{\tau}+i \bk\cdot \bv_\parallel \right) t 
-i\bk\cdot {\bf x}(\bv_\perp, t)\right\} \hat{f_0}\left(\bk, {\cal R}_{\widehat{\bf B}}(-\omega t)\cdot \bv\right)  \]
whereas the function
\begin{equation}
\label{20}
\psi(\bk,\bv,z)
=\frac{1}{\tau }\int_0^{+\infty} dt\, \exp\left\{-\left( z+\frac{1}{\tau}+i \bk\cdot \bv_\parallel \right) t 
-i\bk\cdot {\bf x}(\bv_\perp, t)\right\}
\end{equation}
multiplies the spherical part of the transformed distribution $\tilde{f}$.
Owing to this factorized structure the application of the projection $\mathbb{P}$ to both sides of (\ref{LFTB}) yields a closed equation for $[\mathbb{P}\tilde{f}](\bk,v,z) $ providing the formula
\begin{equation}
\label{21}
[\mathbb{P}\tilde{f}](\bk,\bv,z)=\frac{ \left[\mathbb{P}\mathbb{L}\hat{f_0}\right](\bk,v,z)}
{1-[\mathbb{P}\psi](\bk,v,z)}
\end{equation}
By inserting the solution (\ref{21}) into (\ref{LFTB}) we arrive at the complete solution for the Fourier-Laplace transform of the distribution function
\begin{equation}
\label{LFTsolution}
\tilde{f}(\bk,\bv,z)=\left[\mathbb{L}\hat{f_0}\right](\bk,\bv,z) 
+\frac{\psi(\bk,\bv,z)}{1-[\mathbb{P}\psi](\bk,v,z)}
\times\left[\mathbb{P}\mathbb{L}\hat{f_0}\right](\bk,v,z) 
\end{equation}
The exact evolution in the course of time can be inferred from (\ref{LFTsolution}) by taking the inverse Laplace transformation
\begin{equation}
\label{inverseLT}
\hat{f}(\bk,\bv,t) = \int \frac{dz}{2\pi i}\exp (zt)\tilde{f}(\bk,\bv,z)
\end{equation}
The contour of integration in (\ref{inverseLT}) is a line parallel to the imaginary axis lying to the right of all singularities of the analytic function $\tilde{f}(\bk,\bv,z)$. In order to analyze the modes of the time evolution we have thus to examine the singularities of  functions $\mathbb{L}\hat{f_0}$, $\mathbb{P}\mathbb{L}\hat{f}_0$ and $ \psi $ together with  zeros of the function 
$(1-\mathbb{P}\psi )$ where
\begin{equation}
\label{22}
[\mathbb{P}\psi](\bk,v,z)=
\end{equation}
\[ \frac{1}{\tau }\int_0^{+\infty} dt\, \exp\left\{-\left(z+\frac{1}{\tau}\right)t \right\} 
 \int  \frac{d\widehat{\bv}}{4\pi}\,\exp\left\{-i \bk\cdot \bv_\parallel t-i\bk\cdot {\bf x}
(\bv_\perp, t) \right\} \]
 with
 \begin{equation}
 \label{23}
{\bf x}(\bv_\perp, t) = \frac{1}{\omega}\left[ {\cal R}_{\widehat{\bf B}}\left(\frac{\pi}{2}-\omega t\right)\cdot \bv_\perp - {\cal R}_{\widehat{\bf B}}\left(\frac{\pi}{2}\right)\cdot \bv_\perp\right] 
\end{equation}
(see  definition (\ref{18})).

\section{Separation of time scales: appearance of the hydrodynamic mode of diffusion}

The  formulae   (\ref{19}), (\ref{20}) show that  $\mathbb{L}\hat{f_0}(\bk,\bv,z)$ and $ \psi(\bk,\bv,z)$ are Laplace transforms of  bounded (periodic) functions evaluated at the point
$(z+1/\tau+i \bk\cdot \bv_\parallel  )$ whose real part equals $({\rm Re}z + 1/\tau )$. It follows that all their singularities in the complex $z$-plane are located in the half-plane
${\rm Re}(z)  \le -1/\tau $ (the same holds for $\mathbb{P}\mathbb{L}\hat{f}_0$ ). The modes of evolution related to these singularities correspond thus to a rapid decay dominated by the exponential decrease $\sim \exp (-t/\tau)$. 

A qualitatively different behavior comes from  the singularity occuring when the denominator in (\ref{21})
vanishes. Consider the equation
\begin{equation}
\label{24}
[\mathbb{P}\psi](\bk,v,z)=1
\end{equation}
Its explicit form is conveniently written in the coordinate frame whose $z$-axis is parallel to $\widehat{\bf B}$ and where  vector $\bk$ lies in the plane $xOz$ with coordinates  $(k_\perp,0,k_\parallel )$. We find
\begin{equation}
\label{25}
\frac{1}{\tau }\int_0^{+\infty} dt\, \exp\left\{-\left(z+\frac{1}{\tau}\right)t \right\}\int_0^\pi \frac{ \sin\theta d\theta}{2}\int_0^{2\pi} \frac{d\phi}{2\pi} 
\end{equation}
\[ \times 
\exp\left\{-i t k_\parallel v \cos\theta -i\frac{1}{\omega}\sin\left(\frac{\omega t}{2}\right)2 k_\perp v \sin\theta\sin\phi\right\} = 1 \]
where $ ( v\sin\theta\cos\phi ,v\sin\theta\sin\phi, v\cos\theta )$ are the coordinates of the velocity vector. 
Denoting by $z_{hyd}(\bk )$ the implicit function defined by (\ref{25}) one checks that at $\bk = 0$ there is a unique solution $z_{hyd}(\bk =0)=0$. Upon expanding the integrand in (\ref{25}) in powers of $k_{\perp}$ and $k_{\parallel}$ we find the asymptotic expansion of  $z_{hyd}(\bk )$ for $\bk \to 0$ in the form
\begin{equation}
\label{26}
z_{hyd}(\bk) =-\left[ D_\parallel k_\parallel^2+D_\perp k_\perp^2\right]+o(k^2)
\end{equation}
where the diffusion coefficients $D_\parallel $ and $D_\perp $ coincide with those derived from the Green-Kubo relation (\ref{D}) (see (\ref{Dd})). $o(k^2)$ denotes a term vanishing faster than $k^2$.

The isolated zero at $z=z_{hyd}(\bk)$ of the analytic function $(1-\mathbb{P}\psi )$ corresponds to the hydrodynamic pole in the solution (\ref{LFTsolution}). Indeed, in the vicinity of $z=0$ and for $\bk \to 0$ one finds the asymptotic representation
\begin{equation}
\label{27}
\frac{1}{1-\mathbb{P}\psi (\bk,v,z) } \cong \frac{1}{\tau (z - z_{hyd}(\bk))}
\end{equation}
The evaluation of the inverse Laplace transformation (\ref{inverseLT}) will thus yield the contribution 
\begin{equation}
\label{fhyd}
{\hat f}_{hyd}(\bk,\bv,t) \sim {\rm exp}[-t(D_\parallel k_\parallel^2+D_\perp k_\perp^2)]
\end{equation}
which represents the slow diffusion process spreading the electrons in the position space. Their number density in the Laplace-Fourier representation
(at a given modulus of velocity) follows  directly from (\ref{LFTsolution})  by applying the projector $\mathbb{P}$
\begin{equation}
\label{LFn}
\tilde{n}(\bk,v,z)= 
\frac{4\pi}{1-[\mathbb{P}\psi](\bk,v,z)}
\left[ \mathbb{P}\mathbb{L}\tilde{f_0}\right](\bk,v,z) 
\end{equation}
Here the hydrodynamic zero $z=z_{hyd}$ determines entirely the evolution for long times $t\gg 1/\tau$. 

The magnetic field does not change the kinetic energy of the electrons. (\ref{LFn}) is thus the density of a conserved quantity which implies the appearance of the hydrodynamic mode. The situation here is analogous to that studied in \cite{Hauge70}  in the absence of ${\bf B}$ where the existence of Chapman-Enskog solutions  holding all along the evolution has been proved by construction. In the next section we show how to construct systematically Chapman-Enskog solutions in the presence of the field ${\bf B}$. 

\section{Chapman-Enskog solutions}

The only hydrodynamic field in the dynamics of the Lorentz model is the number density (\ref{LFn}).
In order to investigate the possibility of the existence of  Chapman-Enskog solutions 
depending on time only via $n(\br,v,t)$ we follow here the ideas developed in \cite{Hauge70} where such solutions have been explicitly constructed. In the Fourier representation we thus look for distributions of the form
\bee
\hat{f}(\bk,\bv,t) & = &[\mathbb{P}\hat{f}](\bk,v,t) + [(1 - \mathbb{P})\hat{f}](\bk,\bv,t) \label{28} \\
 & = & \frac{1}{4\pi }\hat{n}(\bk,v,t) + L(\bk,\bv |\hat{n}(.,t)) \nonumber
\eee
If $\hat{f}$ solves the linear Boltzmann equation 
\begin{equation}
\label{FB}
\left[\frac{\partial  }{\partial t} +i\bk\cdot\bv+
\left(\boldsymbol{\omega}\times \bv \right) \cdot\frac{\partial}{\partial \bv}\right]\hat{f}(\bk,\bv,t)=-\frac{1}{\tau}\left[\hat{f}(\bk,\bv,t)-[\mathbb{P}\hat{f}](\bk,v,t)\right]
\end{equation}
then  $\alpha\hat{f}$ is  another solution for any number $\alpha$. It follows that $L(\bk,\bv |\hat{n}(.,t))$ is a linear functional of $\hat{n}$
\begin{equation}
\label{29}
  \alpha L(\bk,\bv |\hat{n}(.,t)) = L(\bk,\bv |\alpha \hat{n}(.,t)) 
\end{equation}  
 Following Hauge \cite{Hauge70} we assume the simplest local linear relation 
 \begin{equation}
 \label{30}
  L(\bk,\bv |\hat{n}(.,t)) = F(\bk,\bv)  \hat{n}(\bk,v,t) 
  \end{equation}
and look for the Chapman-Enskog solutions of the form
\begin{equation}
\label{functionF}
\hat{f}(\bk,\bv,t)=[1+F(\bk,\bv)][\mathbb{P}\hat{f}](\bk,v,t)
\end{equation}
In order to derive a closed equation for the function $F$ we insert (\ref{functionF}) into (\ref{FB}). We get
\begin{equation}
\label{31}
\left[(1+F)\frac{\partial }{\partial t}+i\bk\cdot\bv(1+F)+
\left(\boldsymbol{\omega}\times \bv \right) \cdot\frac{\partial F}{\partial \bv}\right]
[\mathbb{P}\hat{f}](\bk,v,t)=
-\frac{1}{\tau}F\, [\mathbb{P}\hat{f}](\bk,v,t)
\end{equation}
In view of the relations
\begin{equation}
\label{32}
\mathbb{P}[F]=0,
\quad
\mathbb{P}[i\bk\cdot\bv]=0
\quad\textrm{and}\quad
\mathbb{P}\left[\left(\boldsymbol{\omega}\times \bv \right) \cdot\frac{\partial F}{\partial \bv}\right]=0
\end{equation}
the application of the projection $\mathbb{P}$ to (\ref{31}) yields 
\begin{equation}
\label{33}
\left\{\frac{\partial }{\partial t} + [\mathbb{P} (i\bk\cdot\bv ) F]\right\}
[\mathbb{P}\hat{f}](\bk,v,t)=0
\end{equation}
which determines the evolution of the spherical part of the distribution 
\begin{equation}
\label{34}
[\mathbb{P}\hat{f}](\bk,v,t)=
[\mathbb{P}\hat{f}_0](\bk,v)
\exp\{ -t  [\mathbb{P} (i\bk\cdot\bv ) F ](\bk,v) \}
\end{equation}
The time evolution of the complete distribution (\ref{functionF}) has been thus also determined
\begin{equation}
\label{FourierCHE}
\hat{f}(\bk,\bv,t)=[1+F(\bk,\bv)][\mathbb{P}\hat{f}_0](\bk,v)
\exp\{-t  [\mathbb{P}(i\bk\cdot\bv ) F ](\bk,v) \}
\end{equation}
When inserted into (\ref{31}) the equation (\ref{33}) permits to derive the nonlinear equation determining  the function $F$
\begin{equation}
\label{eqF}
\left[\frac{1}{\tau}+\left(\boldsymbol{\omega}\times \bv \right) \cdot\frac{\partial }{\partial \bv}\right]F
=\left\{[\mathbb{P} (i\bk\cdot\bv ) F] -i\bk\cdot\bv \right\} (1+F )
\end{equation}
Notice that by putting $\bk =0$ in (\ref{33}) we find
\begin{equation}
\label{35}
\left\{\frac{\partial }{\partial t} + \lim_{\bk\to 0}[\mathbb{P} (i\bk\cdot\bv ) F]\right\}
[\mathbb{P}\phi](v,t)=0
\end{equation}
where $\phi (\bv ,t)$ is the velocity distribution. (\ref{35}) should coincide with the previously derived condition (\ref{6}) expressing the invariance of the spherically symmetric component of $\phi$. We conclude that
\begin{equation}
\label{36}
\lim_{\bk\to 0}[\mathbb{P} (i\bk\cdot\bv ) F]=0
\end{equation}
Taking then the limit $\bk\to 0$ in (\ref{FourierCHE}) we find that the  velocity distribution $\phi $ does not change in the course of time which is possible only for spherically symmetric initial conditions (see
(\ref{vd})). This property of the Chapman-Enskog solution imposes on $F$ the condition
\begin{equation}
\label{37}
F(\bk=\boldsymbol{0},\bv)=0
\end{equation}
Indeed, as the relation (\ref{functionF}) must hold at any time, in particular at $t=0$
\begin{equation}
\label{initialCE}
\hat{f}(\bk,\bv,0)=[1+F(\bk,\bv)][\mathbb{P}\hat{f}](\bk,v,0),
\end{equation}
and the restriction  of the class of initial conditions to those whose velocity distributions are spherically symmetric enforces (\ref{37}). 
 
 Coming back to the Laplace-Fourier transform $\tilde{f}$ we get from (\ref{FourierCHE}) the formula
 \begin{equation}
 \label{38}
 \tilde{f}(\bk,\bv,z)
=\left[1+F(\bk,\bv)\right][\mathbb{P}\hat{f}_0](\bk,v)
\frac{1}{z+ [\mathbb{P}(i\bk\cdot\bv )F ](\bk,v)}
 \end{equation}
 The Chapman-Enskog distribution $\tilde{f}(\bk,\bv,z)$ has thus a single hydrodynamic pole at 
 \begin{equation}
 \label{pole}
 z=z_{hyd}(\bk)=-[\mathbb{P}(i\bk\cdot\bv )F ](\bk,v)
 \end{equation}
which determines entirely the evolution in the course of time. 
    
    A subtle point should be noted here.
    Owing to the presence of spatial gradients the Chapman-Enskog distribution is not spherically symmetric as function of ${\bf v}$ all along the evolution towards equilibrium. However,  as a consequence of (\ref{37}) its ${\bf k}={\mathbf 0}$ component  possesses the spherical symmetry at any moment of time.   
  
The power series representation of the function $F$  can be constructed in a systematic way in terms of convenient variables (see (\ref{25}))
\bee
\label{39}
s_1& = & \bk_\perp\cdot\bv_\perp=k_\perp v \sin\theta\cos\phi \\
s_2 & = & \left(\bk_\perp\times\bv_\perp\right)\cdot\widehat{\bf B}=k_\perp v \sin\theta\sin\phi 
\nonumber \\
s_3 & = & \bk_\parallel\cdot\bv_\parallel=k_\parallel v \cos\theta \nonumber
\eee
 We write 
  \begin{equation}
  \label{series}
  F=F^{(1)}+F^{(2)}+\ldots
  \end{equation}
  where $F^{(n)}$ is the $n$-th order polynomial in variables $(s_1,s_2,s_3)$. 
  The action of the operator 
  ${\cal L}=\left(\boldsymbol{\omega}\times \bv \right) \cdot\partial /\partial \bv $ 
  appearing in (\ref{eqF}) on these variables is particularly simple
  \begin{equation}
\label{40}
{\cal L}s_1=-\omega  s_2,\quad{\cal L}s_2=\omega s_1\quad\textrm{and}\quad {\cal L}s_3=0
\end{equation}
which greatly facilitates the calculations. The nonlinear equation (\ref{eqF}) implies the chain of linear  equations 
\begin{equation}
\label{order1}
\left[\frac{1}{\tau}+\left(\boldsymbol{\omega}\times \bv \right) \cdot\frac{\partial }{\partial \bv}\right]F^{(1)}(\bk,v)
=-i\bk\cdot\bv 
\end{equation}
 \begin{equation}
\label{order2}
\left[\frac{1}{\tau}+\left(\boldsymbol{\omega}\times \bv \right) \cdot\frac{\partial }{\partial \bv}\right]F^{(2)}(\bk,v)
= [\mathbb{P} (i\bk\cdot\bv) F^{(1)}](\bk,v)-i\bk\cdot\bv  F^{(1)}
\end{equation}
\[   ...........................\]
which can be consecutively solved. For the first two terms of the expansion one finds
\begin{equation}
\label{F1}
F^{(1)}(\bk,\bv)=-i\tau\left[s_3+\frac{1}{1+\omega^2\tau^2}\left(s_1+\omega\tau s_2\right)\right]
\end{equation}
\begin{eqnarray}
\label{F2}
F^{(2)}(\bk,\bv)&=&
\tau\left[D_\parallel k_\parallel^2+D_\perp k_\perp^2\right]
-\tau^2s_3^2
\\
&-&\tau^2\frac{1}{(1+\omega^2\tau^2)^2}
\left[2s_1s_3+\omega\tau(3+\omega^2\tau^2)s_2s_3\right]
\nonumber
\\
&-&
\tau^2\frac{1}{(1+\omega^2\tau^2)(1+4\omega^2\tau^2)}
\left[(1+\omega^2\tau^2)s_1^2+3\omega\tau s_1s_2+3\omega^2\tau^2s_2^2\right]
\nonumber
\end{eqnarray}

According to (\ref{pole}) the second order term  in the expansion of the hydrodynamic pole  is given by
 \begin{equation}
\label{zhyd2} 
z^{(2)}_{hyd}=- [\mathbb{P} (i\bk\cdot\bv ) F^{(1)}](\bk,v)=
-\left[D_\parallel k_\parallel^2+D_\perp k_\perp^2\right]
\end{equation}
which reproduces the already derived formula (\ref{26}). For symmetry reasons any term $F^{(2n)}$ in (\ref{series}) is a polynomial of even order in the variables $s_1$, $s_2$ and $s_3$. Therefore, by virtue of definitions (\ref{39}), $F^{(2n)}$  is invariant by the change $\bv\to -\bv$ and $[\mathbb{P} (i\bk\cdot\bv ) F^{(2n)}](\bk,v)=0$ : the small $\bk$ expansion of the hydrodynamic pole contains only terms of even orders in $k$.

Notice that (\ref{33}) is in fact the equation for the number density
\begin{equation}
\label{41}
\left[\frac{\partial }{\partial t} + [\mathbb{P} (i\bk\cdot\bv )F ]\right]\hat{n}(\bk,v,t)=0
\end{equation}
 In the small $\bk$ limit we can use here the first order approximation $F^{(1)}$.
 In the position space this gives the diffusion equation
 \begin{equation}
\label{Deq}
\frac{\partial }{\partial t} n(\br,v,t)= \left[D_\parallel \Delta_\parallel+D_\perp \Delta_\perp\right]n(\br,v,t)
\end{equation}
$\Delta_\parallel $ and $ \Delta_\perp $ are the Laplace operators corresponding to space variables
conjugate to the wave vectors $\bk_{\parallel}$ and $\bk_{\perp}$, respectively.
 
In principle, one could continue the evaluation of higher order terms going beyond the classical diffusion equation (\ref{Deq}). In fact, the determination of the $n$-th order polynomial
$F^{(n)}$ in the series (\ref{series})  is straightforward once one knows all  polynomials of lower rank.
 To prove the existence of Chapman-Enskog solutions there remains the open question of convergence of expansion (\ref{series}).

\section{Discussion and comments }

  The exact solution (\ref{LFTsolution}) of the Boltzmann equation for the three dimensional Lorentz model in the presence of a constant and uniform magnetic field allows to  describe precisely the dynamical evolution of the system much like in the previously studied case {\cite{Hauge70}} of free motion between collisions.
The picture which emerges from our analysis bears strong resemblance to the case with ${\bf B}=0$. This is clearly related to the fact that by taking the limit $\omega = eB/m \to 0$ in  various predictions following from (\ref{LFTsolution})  we recover the results of  \cite{Hauge70}.
We could thus reveal the time scale separation between the rapid exponential decay (\ref{vd}) of the deviation from the spherical symmetry in the velocity distribution and the slow  diffusive spreading  in the position space due to the hydrodynamic pole (\ref{26}). This was achieved by showing that  the Laplace-Fourier transform $\tilde{f}(\bk,\bv,z)$ contained a single pole 
approaching $z=0$ for $\bk\to 0$ whereas the rest of the singularities of $\tilde{f}$ remained within  the half-plane ${\rm Re}z \leq -1/\tau $. 
We have also checked that the Green-Kubo formula (\ref{D}) gave the exact value for the diffusion coefficients
$D_{\perp},\, D_{\parallel} $ by identifying them independently in the small wave vector expansion of the
 hydrodynamic pole (\ref{26}).
 Of course, there is here a new effect compared to \cite{Hauge70}. The magnetic field introduces an anisotropy into the diffusion process absent in the field free case.
 For the diffusion in the direction perpendicular to ${\bf B}$ the Lorentz model predicts the decay 
 $D_{\perp} \sim (\omega\tau)^{-2}$ in the limit of strong fields $\omega\tau\gg 1$. The diffusion parallel to ${\bf B}$ looks the same as in  \cite{Hauge70}. 

There is one point where the presence of the magnetic field makes the problem much more difficult than in the case of ${\bf B}=0$. It concerns the possibility of solving the nonlinear equation (\ref{eqF}), necessary for the explicit construction of the Chapman-Enskog solutions. In the limit $\omega\to 0$ (\ref{eqF}) takes the form
\begin{equation}
\label{eqF0}
\frac{1}{\tau}F
=\left\{ [\mathbb{P} (i\bk\cdot\bv ) F] -i\bk\cdot\bv \right\} (1+F )
\end{equation}
and can be readily solved yielding
\begin{equation}
\label{F0}
F(\bk,\bv)|_{{\bf B}=0} = \frac{1}{kv\tau\cot(kv\tau ) + i\tau \bk\cdot\bv} -1
\end{equation}
It is quite remarkable that the solution of the nonlinear equation (\ref{eqF0}) is so easily obtained. It permitted in \cite{Hauge70} to provide the proof of the existence of the Chapman-Enskog solutions by explicit construction.
This was however not the case with our equation (\ref{eqF}). All we were able to do was to find the way to a systematic expansion of $F(\bk,\bv)$ in a power series of the variables 
$k_{\perp},\, k_{\parallel}$. The first two terms (\ref{F1}) and (\ref{F2}) of the expansion have been exhibited in the text. They become $(-i \tau\bk\cdot\bv)$ and $[-(\tau \bk\cdot\bv)^{2}+(\tau k v)^{2}/3]$, respectively, in the $\omega\to 0$ limit reproducing the first two terms in the  expansion of (\ref{F0}). The still open question is the convergence of the complete series (\ref{series}). 
   
   Let us finally remark, that the further generalization consisting in including also an electric field would require a modification in the dynamics of the Lorentz model, as keeping elastic collisions with fixed scatterers excludes the existence of a stationary state  because of the unbounded absorption of energy \cite{PW79}.

\medskip   
{\bf Acknowledgments}

 J. P. greatly 
acknowledges the hospitality at the Laboratoire de Physique Th\'eorique de
l'Universit\'e Paris-Sud.

\end{document}